# Visible Photoluminescence from Cubic (3C) Silicon Carbide Microdisks Coupled to High Quality Whispering Gallery Modes

Marina Radulaski,*,† Thomas M. Babinec,† Kai Müller,† Konstantinos G. Lagoudakis,† Jingyuan Linda Zhang,† Sonia Buckley,† Yousif A. Kelaita,† Kassem Alassaad,‡ Gabriel Ferro‡ and Jelena Vučković†

†E. L. Ginzton Laboratory, Stanford University, Stanford, California 94305, USA

‡Laboratorie des Multimateriaux et Interfaces, Universite de Lyon, 69622 Villeurbanne Cedex, France

**Abstract**: We present the design, fabrication and characterization of cubic (3C) silicon carbide microdisk resonators with high quality factor modes at visible and near infrared wavelengths (600 - 950 nm). Whispering gallery modes with quality factors as high as 2,300 and corresponding mode volumes V ~ 2 × $(\lambda/n)^3$ are measured using laser scanning confocal microscopy at room temperature. We obtain excellent correspondence between transverse-magnetic (TM) and transverse-electric (TE) polarized resonances simulated using Finite Difference Time Domain (FDTD) method and those observed in experiment. These structures based on ensembles of optically active impurities in 3C-SiC resonators could play an important role in diverse applications of nonlinear and quantum photonics, including low power optical switching and quantum memories.

## Introduction

Photonic microcavities provide an integrated architecture for the confinement of optical fields into small volumes with long photon lifetimes ($\tau$).[1] The ratio of the cavity quality factor ($Q \sim \tau$) to overall mode volume (*V*) provides a figure of merit for applications ranging from spontaneous emission enhancement of embedded optical emitters,[2] to non-classical light generation[3] and single photon nonlinear optics.[4] High cavity quality factors and wavelength-sized microdisk resonators are one cavity system that is well suited for all of these applications.

Many optical emitters operate in the visible wavelength regime and there have recently been several demonstrations of optical microcavities operating at these wavelengths, in material systems such as gallium phosphide,[5] silicon nitride,[6] titanium dioxide[7] and diamond.[8] Silicon carbide, on the other hand, has recently emerged as an attractive alternative material system offering a unique combination of wide bandgap and transparency window, significant second order ($\chi^{(2)}$) nonlinearity,[9] optically active impurities (color centers) possessing addressable spin,[10] and potential for spin quantum memories possessing long coherence times via the principle of the 'semiconductor vacuum'.[11,12]

Silicon carbide photonics is itself a broad field due to the many (~250) different stacking sequences (polytypes) of silicon carbide. Three of these, cubic 3C and hexagonal 4H and 6H, have been actively present in materials, MEMS and photonics research. Despite the similarity of many of their properties, such as refractive index, magnitude of nonlinear optical coefficients and etch rate, these polytypes manifest strikingly different lattice symmetries. As a result, only the cubic polytype is able to be grown directly on to a silicon substrate. This makes 3C-SiC a favorable polytype for applications where direct integration with silicon is of interest, as well as for fabrication of planar devices based on films on a sacrificial layer. Obtaining slabs from bulk hexagonal polytypes is usually done by the Smart-Cut process,[13] which may cause surface roughness, as well as generate undesired impurities in the sample. In contrast, heteroepitaxial growth of cubic silicon carbide on silicon has superior sample properties.

One of the basic building blocks of an on-chip photonics platform is a high quality factor microcavity. There have been successful demonstrations of photonic crystal cavities in 6H-SiC from visible to telecommunication wavelengths,[14] as well as their applications for nonlinear frequency conversion.[9] Similarly, electrochemically etched microdisks in epitaxial p-doped layers of 4H-SiC have been shown in the visible.[15] Demonstrations in 3C-SiC have, so far, been limited to infrared wavelengths in photonic crystals cavities[16,17] and microdisks.[18] This includes coupling photonic crystal cavities to Ky5 color center near 1.1 µm wavelength.[17]

High quality microresonators in cubic silicon carbide at visible wavelengths would have several applications. For example, coupling intense optical fields to the emission of embedded optically active silicon vacancy (at 645 nm wavelength, the origin of this impurity being recently referred as carbon anti-site vacancy[19]) and silicon divacancy[20] (629 nm) are very interesting for applications of cavity quantum electrodynamics (cQED) such as non-classical light generation,[21] low-power optical switching[22] and spectroscopy.[23] In particular, due to the atomic size of these impurities (color centers), such a platform would allow experiments in a challenging regime of fundamental physics, multi-emitter cQED, whereby $N$ identical emitters could couple to a common mode in order enhance the light-matter coupling rate $g$ to $g\sqrt{N}$. This regime has been inaccessible in solid state systems based on quantum dots, such as InAs/GaAs, due to the mesoscopic size of the emitters and practical challenges associated with growing multiple identical quantum dots.[24] Our goal is to demonstrate a multi-emitter cavity QED system on a semiconductor platform, akin to the multi-emitter atomic cavity QED system,[25] and featuring a ladder of dressed states, long coherence times, and room temperature operation necessary for many practical applications including optical switching, nonclassical light generation, and quantum memories. We note that despite remarkable progress with semiconductor,[26] organic[27] and erbium[28] microcavity polaritons, the aforementioned features cannot be achieved in those platforms.

Here, we demonstrate coupling of visible to infrared intrinsic emission from 3C-SiC to high quality microdisks fabricated in this material with 1.7 – 2.5 µm diameter and supported by silicon pedestals. Pushing the fabrication limits of 3C-SiC microdisk resonators down to sub 2 µm diameters, we obtain a new generation of structures capable of confining visible light within high quality resonances. We apply confocal photoluminescence techniques to characterize the microdisks at room temperature, and detect low-level intrinsic substrate luminescence coupling to whispering gallery modes. We measure quality factors as high as 2,300, with calculated mode volumes $[1.8, 4.6] \times (\lambda/n)^3$. In addition to accessing the new set of wavelengths, these structures also benefit from reduced fabrication complexity compared to recent demonstrations of photonic crystals in 3C-SiC.[16,17] The size and position control in our samples are a new feature, compared to the microdisks of similar dimension in hexagonal polytypes, which use dispersed microspheres of variable size to define the etch mask.[16] Finally, our structures are an order of magnitude smaller than in previous 3C-SiC microdisk demonstration.[18]

**Fabrication of 3C-SiC microdisks**

Figure 1 shows the process flow we use to fabricate 3C-SiC microdisks. We begin with a 3C-SiC film of approximate 210 nm thickness, grown on Si by a standard two-step chemical vapor deposition technique.[29] In our previous demonstration of telecom wavelength photonic crystal nanocavities in the same material,[16] we used positive tone resist to define holes in the pattern. For the microdisks examined in this work, we have used the negative tone e-beam resist Microposit ma-N 2403, followed by patterning with a 100 keV electron beam tool. Silicon carbide was etched in HBr-$Cl_2$ plasma, while the underlying silicon was partially removed in $XeF_2$ gas phase etcher (for details, see Methods).

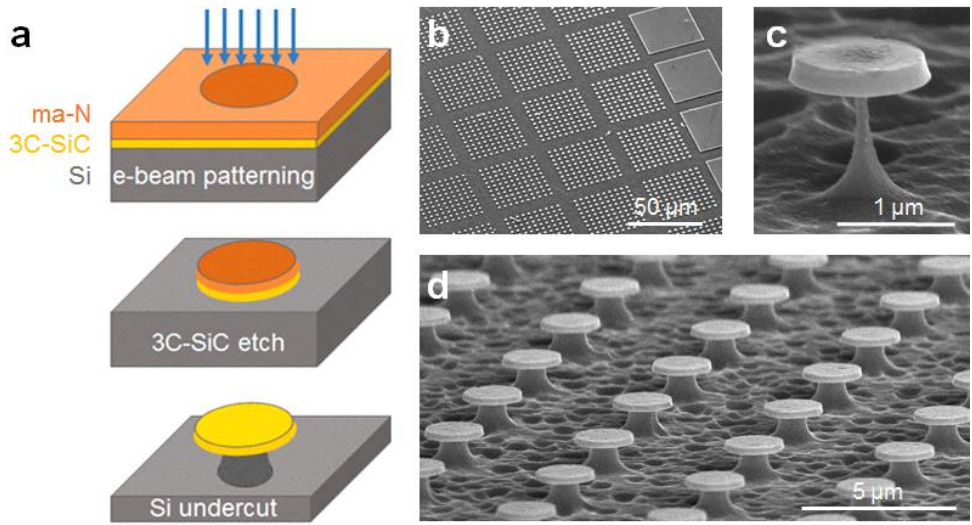

**Figure 1. a** 3C-SiC microdisk fabrication process flow. **b** Fabricated 10 x 10 arrays of microdisks and 50 µm x 50 µm square structures. **c** The smallest fabricated microdisk with diameter 1.1 µm on a 100 nm wide pedestal. **d** Array of robust 1.9 µm diameter SiC microdisks on 900 nm wide Si pedestals.

Microdisks were fabricated with diameters varying between 1.1 µm and 2.5 µm, in 10 x 10 arrays, alongside 50 µm x 50 µm square structures, later used as a reference in photoluminescence measurements (Figure 1b). Additional, pre-optical measurement, cleaning process in 3:1 piranha solution was performed to remove any remaining organic residue. Microdisks of diameter 1.7 µm and higher (Figure 3d) were robust, with a satisfactory etch profile and undercut area that can support first order radial whispering gallery modes. Figure 1c shows the smallest fabricated microdisk of 1.1 µm diameter on a 100 nm thin pedestal (although this one could not be characterized, as it turned out too fragile and fell off the pedestal during the piranha solution cleaning).

**Simulated whispering gallery modes**

We used the Finite Difference Time Domain (FDTD) method to simulate whispering gallery modes supported in the microdisks. Figure 2a shows quality factors and wavelengths of simulated modes supported in a SiC microdisk (index $n = 2.6$) with diameter 1.7 µm and thickness 210 nm, standing on a Si ($n = 3.77$) cylindrical pedestal with diameter 700 nm and height 500 nm. Material absorption was not included in the model, therefore the quality factors presented have purely radiative character ($Q_{rad}$), which exponentially decays with mode wavelength. For wavelengths shorter than 700 nm, multiple mode overlap introduced error in the extraction of $Q_{rad}$, therefore the plotted values represent its lower boundary.

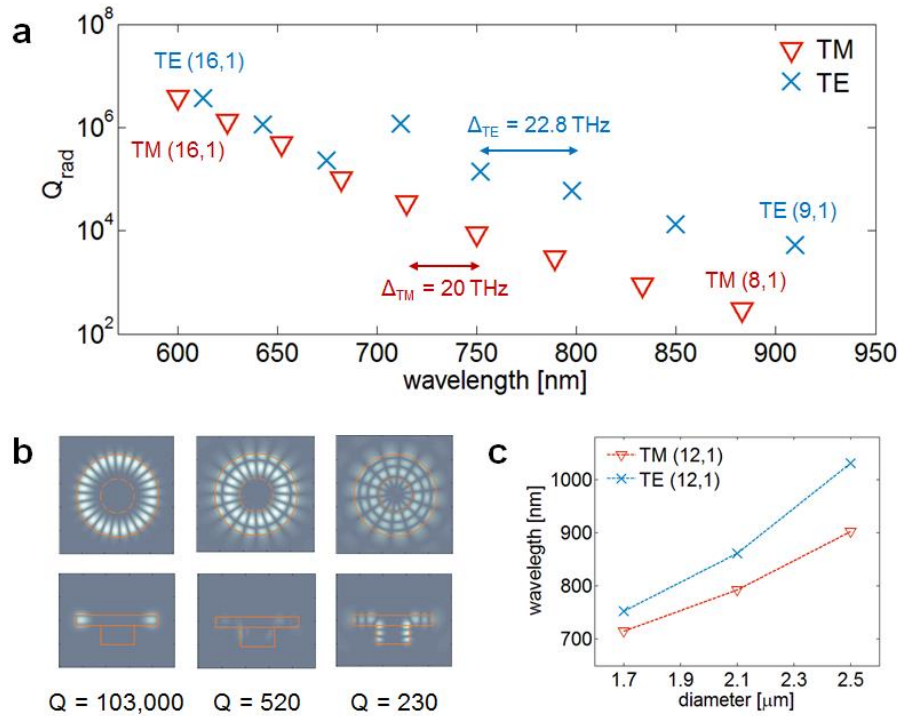

**Figure 2.** FDTD simulations of microdisks. **a** Radiative quality factor of TM and TE first radial order ($n = 1$) modes supported in a 1.7 µm diameter disk on a 700 nm pedestal; free spectral range between subsequent modes is marked by arrows. **b** $|E_z|$ component of electric field in plane of the microdisk and in vertical plane of the structure, for whispering gallery TM modes (13,1), (10,2), (7,3) at wavelengths 682 nm, 675 nm, 662 nm, respectively, with quality factors noted beneath. **c** Whispering gallery (12,1) mode wavelength scaling with disk size for TM and TE polarization.

We identified whispering gallery modes $(m,n)$ with radial mode $n = 1$ and azimuthal (angular) modes $8 \leq m \leq 16$ for transverse-magnetic polarization (TM, electric field orthogonal to the plane of the disk) and $9 \leq m \leq 16$ for transverse-electric polarization (TE, electric field in the plane of the disk). The quality factors were as high as $Q = 3.7 \times 10^6$, and the mode volumes were in the range $V \epsilon [1.8, 4.6] \times \left(\frac{\lambda}{n}\right)^3$. Neighboring $(m,1)$ modes were frequency $(f = c/\lambda)$ separated by free spectral range $\Delta_{TM}$ = 20.0 THz for TM polarization, and $\Delta_{TE}$ = 22.8 THz for TE polarization. For 1.9 μm diameter disks, these values were $\Delta_{TM}$ = 18.1 THz and $\Delta_{TE}$ = 19.8 THz, while for 2.5 μm diameter disks, they were $\Delta_{TM}$ = 13.2 THz and $\Delta_{TE}$ = 15.5 THz, with uncertainty of 3%. The polarization dependence of the free spectral range is caused by the different mode confinement in TM and TE modes, where the modal region outside of the disk volume lowers the effective index of the mode, and therefore changes its group velocity.

Higher order radial modes were not supported with high quality factor, due to their leakage into the pedestal. Figure 2b shows examples of 1st, 2nd and 3rd order radial modes at nearby wavelengths and illustrate the dramatic three orders of magnitude deterioration of quality factor to $Q \propto 10^2$ for higher order modes. To illustrate the tuning of the mode wavelength with the size of the disk, Figure 2c shows whispering gallery (12,1) modes for both TM and TE polarization and the range of fabricated microdisk diameters. We observe a steeper scaling for TE polarized modes, indicating that for the same set of wavelengths larger microdisks support higher angular order modes, and that TE modes have higher free spectral range than TM modes.

**Characterization of visible resonances in 3C-SiC microdisks**

In this experiment, we use a 2D scanning photoluminescence (PL) setup shown on Figure 3a to characterize the 3C-SiC sample and observe microdisk resonances at room temeprature. Green pump laser is reflected off a 2-axis scanning voice-coil mirror, whose deflection is mapped on to the sample by a 4-f optical configuration and a high numerical apperture lens (for details, see Methods). Filtered signal is collected in the red part of the spectrum. An avalanche photodiode (APD) detects the emitted photons in order to build up a scanned image, while a spectrometer with a a liquid nitrogen cooled Si CCD is used to spectrally resolve the fluorescence.

In order to investigate any low-level fluorescence from the sample we measure the PL spectrum of the chip at different locations on the sample. A scanning confocal microscope image corresponding to our 3C-SiC microdisk structures (Figure 3b) is shown in Figure 3c. We observe an increase in PL signal from the freestanding 3C-SiC film relative to the background with a broad peak at 700 nm. This signal was used to characterize the designed microdisk resonances in an active measurement.

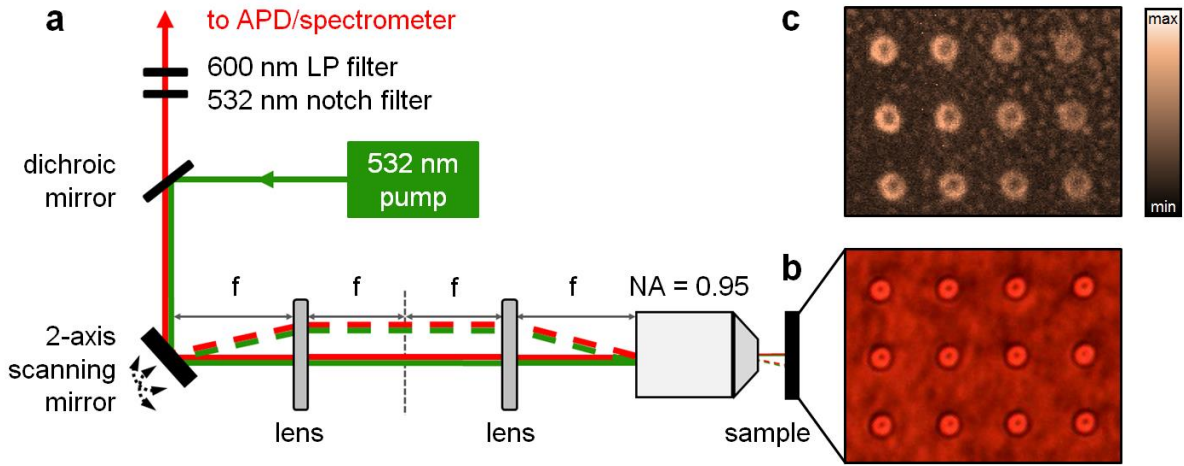

**Figure 3. a** 2D green/red confocal photoluminescence imaging setup used to optically address 3C-SiC microdisks. **b** Optical micrograph of the microdisks sample. **c** 2D photoluminescence scan showing increased signal in the undercut regions of microdisks; maximum signal corresponds to several hundred counts for 0.025 s integration.

We studied 100 keV e-beam irradiation as a possible source of the broad 700 nm PL peak, that may have occurred during SiC irradiation through the negative tone resist during e-beam lithography. As a control sample, we fabricated nonirradiated suspended membranes using positive resist. The spectra in the two approaches were identical at both room temperatures and at 10 K, therefore we ruled out the option of the signal being induced by e-beam irradiation. The likely cause of this PL is the G-band emission from 3C-SiC donor-acceptor recombination in defects close to Si interface,[20] or alternatively emisssion due to fabrication and processing (etching effects or residual organics).

Figure 4a shows PL spectra for microdisk resonators of diameter 1.7 μm, 1.9 μm and 2.5 μm. Resonances with quality factor from 500 to 2,300 are visible along the broad PL peak, with maxima shown in Figure 4b. The obtained spectra were closely reproducible across microdisks in the same array, with wavelength shifts of less than 1 nm. Red triangles and blue crosses mark sets of resonances with frequency spacings closely resembling free spectral range values simulated for TM and TE polarization, respectively. We conclude these mode spacings indeed represent free spectral range in sets of TM and TE polarized resonances in experimental spectra. For 1.7 μm diameter disks, these values were found to be $\Delta_{TM}$ = 20.6 THz and $\Delta_{TE}$ = 22.9 THz across the spectrum. For 1.9 μm they were $\Delta_{TM}$ = 17.4 THz and $\Delta_{TE}$ = 19.8 THz, while for 2.5 μm diameter disks, they $\Delta_{TM}$ = 13.2 THz and were $\Delta_{TE}$ = 15 THz, with 5% uncertainty. A comparison between the measured and simulated free spectral range yields 95% agreement and is shown in Figure 5a. We note that values in simulation and experiment were obtained independently of one another.

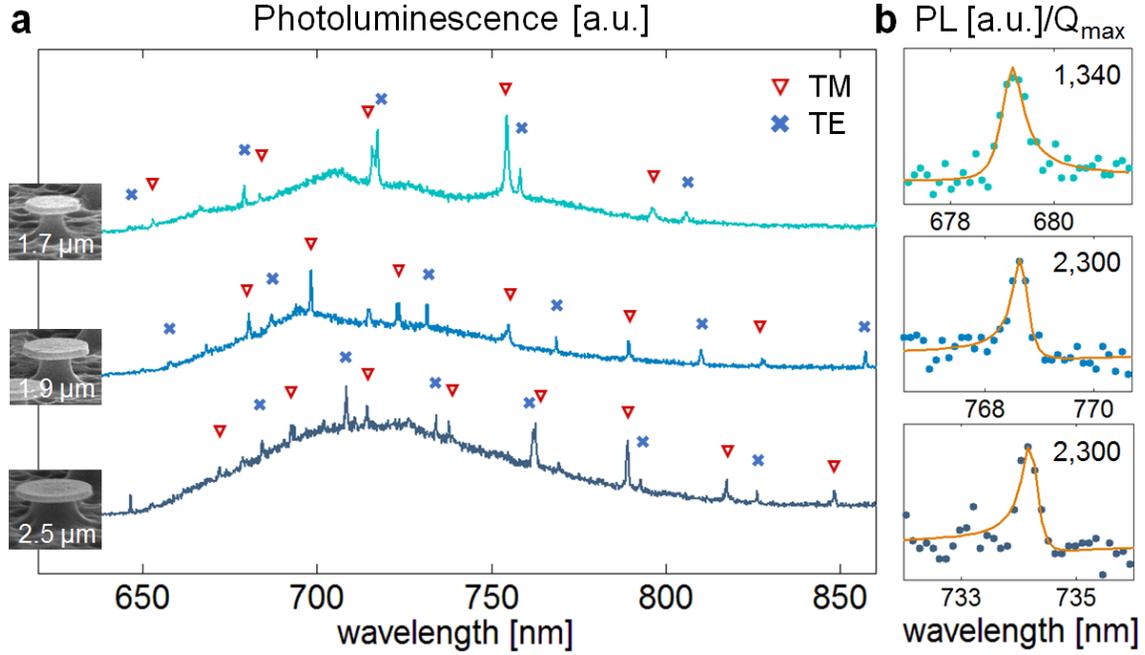

**Figure 4. a** Room temperature photoluminescence signal (linear scale) collected from corresponding microdisks with diameters 1.7 μm, 1.9 μm and 2.5 μm, on 700 nm, 900 nm and 1.5 μm wide pedestals, respectively; red triangles and blue crosses mark identified TM and TE mode series. **b** Maximal quality factor resonance for each disk size.

Next, we tried to deduce the ($m$, $n$) orders of the characterized whispering gallery modes. As discussed under simulation results, the radial mode is determined to be $n = 1$ for all resonances in the series, due to the value of their quality factors. To determine the angular order $m$ of the experimentally obtained TM and TE series of resonances, we found the closest wavelength match to the simulation results. Figure 5b shows the best fit for 1.7 μm diameter microdisks, where the shortest wavelength experimental resonances correspond to modes with $m_{TM} = 14$ and $m_{TE} = 15$. For 1.9 μm diameter disks the best fit was obtained for $m_{TM} = 15$ and $m_{TE} = 16$, while for 2.5 μm diameter disks it was $m_{TM} = 20$ and $m_{TE} = 21$. As expected, larger disks have higher angular order modes at the same set of wavelengths. We believe we have successfully identified TM and TE polarized modes in the microdisks, which is of special interest for coupling to the impurities of known orientation in the lattice.

The possible sources of quality factor reduction in experiment, compared to the simulation, may be the imperfect etch profile, absorption of silicon carbide at the lattice mismatched interface with silicon, or absorption in the silicon pedestal whose band gap is below the energy of the mode. To study the contribution of absorption in the pedestal, we ran additional FDTD simulations to analyze the absorptive loss, assuming the extinction coefficient value of $k = 0.012$. Figure 5c compares the values of calculated radiative and absorptive quality factor, alongside with the experimentally obtained value, for a sample mode in a 1.7 μm diameter disk. We conclude that the pedestal absorption is not the leading source of loss in the system.

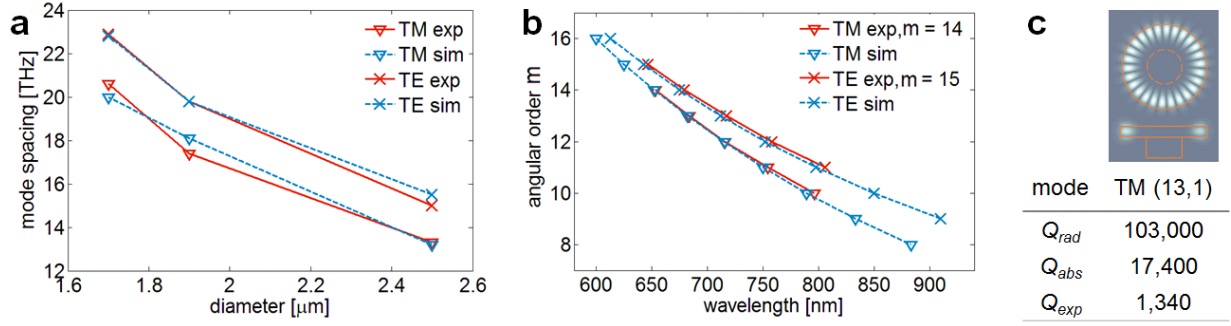

**Figure 5. a** Comparison between simulation results and experimentally obtained free spectral range for TE and TM polarizations. **b** Comparison between simulation results and experimentally characterized resonances in 1.7 μm diameter microdisks, assuming the angular modes $m_{TM} = 14$ and $m_{TE} = 15$ of the shortest wavelength peaks. **c** Comparison between radiative, absorptive and experimentally measured quality factor for TM (13,1) mode in 1.7 μm diameter microdisks.

**Conclusion**

We presented the high quality visible and near-IR resonances coupled to the intrinsic 3C-SiC luminescence in 1.7 – 2.5 μm diameter 3C-SiC microdisks supported by a silicon pedestal. The origin of the luminescence is still under investigation, but possible origins include donor-acceptor G-band recombination. Pushing the fabrication limits of 3C-SiC microdisk resonators down to the record diameters (sub 2 μm), we obtained a new generation of structures capable of confining visible light in high quality resonances previously unobserved in this material. Using confocal photoluminescence techniques, we characterized quality factors as high as 2,300 in low-level intrinsic photoluminescence with calculated mode volumes $[1.8, 4.6] \times (\lambda/n)^3$. Through a theoretical analysis, we identified TM and TE polarized modes, and deduced their whispering gallery radial and angular order. The successful demonstration of SiC microdisk resonators with high Q factors at room temperature in the visible paves the way for the exploration of cavity-QED effects for room temperature switching and nonlinear frequency conversion. Some of the approaches involve doping the 3C-SiC films with emitters such as the di-vacancy, silicon-vacancy or G-band emitters, as well as coupling these microresonators to impurities in other nanocrystalline materials such as vacancies in ZnO or silicon-vacancies in diamond.

**Methods**

**Fabrication** Microdisk fabrication was based on the use of negative tone e-beam resist Microposit ma-N 2403, followed by patterning with a 100 keV electron beam tool (JEOL JBX 6300) with a 375 μC/cm² dose. Development was performed with standard Microposit MF-319 developer. To increase directionality of our SiC etch, in this work we transferred the HBr-Cl$_2$ process to a TCP/RIE Lam Research (9400 TCP Poly Etcher) tool, which provides a uniform, high density transformer coupled plasma. Acetone sonication was used to remove the residual resist from the structures, after which the underlying silicon was etched for three 20 second cycles at 800 mTorr

XeF$_2$ and 50 Torr N$_2$ gas pressures in Xactix e-1 etcher, to release the outer 500 nm along the microdisk radius.

**Confocal photoluminescence measurement** A strong (3 mW) green pump laser is reflected off of a dichroic mirror (Semrock LM01-552-25) towards a 2-axis scanning voice-coil mirror (Newport model FSM-300-01). From here, the deflection of the green pump beam is imaged onto a high numerical aperture objective lens (Olympus SPLAN 100X, NA = 0.95) in a 4-f imaging configuration. Angular deviation of the scanning mirror is converted to real space deflection on the chip. Emitted fluorescence is collected back through the objective, then transmitted through the dichroic mirror and collected in a single mode fiber acting as a confocal pinhole. An additional 532 nm notch filter rejects reflected green pump laser light and a 600 nm long pass filter assures collection of the signal only in the red part of the spectrum.

**Acknowledgements**

This work is supported by NSF DMR Grant Number 1426028. MR and JLZ are supported by Stanford Graduate Fellowship. TMB is supported by the Nanoscale and Quantum Science and Engineering Postdoctoral Fellowship and the AFOSR MURI for Quantum Metaphotonics. KGL acknowledges support by the Swiss National Science Foundation. KM acknowledges support from the Alexander von Humboldt foundation. We thank J Provine, Jeff Hill and Michael Armen for helpful discussions, fabrication and experiment assistance. We thank Alexander Piggott and Jan Petykiewicz for help with FDTD simulation. This work was performed in part at the Stanford Nanofabrication Facility of NNIN supported by the National Science Foundation under Grant No. ECS-9731293.